\documentclass[aps,prb,reprint,superscriptaddress,amssymb,amsmath]{revtex4-1}
\usepackage{graphicx}% Include figure files
\usepackage{dcolumn}% Align table columns on decimal point
\usepackage{bm, color}
\usepackage{amsfonts}
\usepackage[none]{hyphenat} % avoid the hyphen "-" that splits a word in the end of a line.
\usepackage{bbm}
\usepackage{braket}
\usepackage{tabularx}
\usepackage{ulem}

\begin{document}

\title{Topological surface Fermi arcs in the magnetic Weyl semimetal Co$_3$Sn$_2$S$_2$}

\author{Qiunan Xu}
\affiliation{Max Planck Institute for Chemical Physics of Solids, Dresden 01187, Germany}
\author{Enke Liu}
\affiliation{Max Planck Institute for Chemical Physics of Solids, Dresden 01187, Germany}
\affiliation{Institute of Physics, Chinese Academy of Sciences, Beijing 100190, China}
\author{Wujun Shi}
\affiliation{Max Planck Institute for Chemical Physics of Solids, Dresden 01187, Germany}
\affiliation{School of Physical Science and Technology, ShanghaiTech University, Shanghai 200031, China}
\author{Lukas Muechler}
\affiliation{Department of Chemistry, Princeton University, Princeton, New Jersey 08544, USA}
\author{Jacob Gayles}
\author{Claudia Felser}
\author{Yan Sun}
\email{ysun@cpfs.mpg.de}
\affiliation{Max Planck Institute for Chemical Physics of Solids, Dresden 01187, Germany}

\date{\today}

%%%%%%%%%%%%%%%%%%  abstract %%%%%%%%%%%%%%%%%%%
\begin{abstract}
Very recently, the half-metallic compound Co$_3$Sn$_2$S$_2$ was predicted to be a magnetic WSM with Weyl points only 60 meV above the Fermi level ($E_F$). 
Owing to the low charge carrier density and large Berry curvature induced, 
Co$_3$Sn$_2$S$_2$ possesses both a large anomalous Hall conductivity (AHC) and a large anomalous Hall angle (AHA), 
which provide strong evidence for the existence of Weyl points in Co$_3$Sn$_2$S$_2$. 
In this work, we theoretically studied the surface topological feature of Co$_3$Sn$_2$S$_2$ and its counterpart
Co$_3$Sn$_2$Se$_2$. By cleaving the sample at the weak Sn--S/Se bonds, one can achieve two different surfaces terminated with Sn and S/Se atoms, respectively. 
The resulting Fermi arc related states can range from the energy of the Weyl points to $E_F$--0.1 eV in the Sn-terminated surface.
Therefore, it should be possible to observe the Fermi arcs in angle-resolved photoemission spectroscopy (ARPES) measurements. 
Furthermore, in order to simulate quasiparticle interference (QPI) in scanning tunneling microscopy (STM) measurements, 
we also calculated the joint density of states (JDOS) for both terminals. 
This work would be helpful for a comprehensive understanding of the topological properties of these two magnetic WSMs 
and further ARPES and STM measurements.
\end{abstract}
%%%%%%%%%%%%%%%%%%  abstract %%%%%%%%%%%%%%%%%%%

%\pacs{71.20.-b, 73.20.-r, 73.43.-t}
\maketitle

%%%%%%%%%%%%%%%%%%%%%%%%%%%%%%%%%%%%%%%%%%%%%%%%%%%%%%%%%%%%%%%%%%%%%%%
\section{Introduction}
Following the discovery of topological insulators (TIs)~\cite{qi2011RMP, Hasan:2010ku}, 
topological band theory was successfully applied to metals, 
revealing a number of different topological semi-metallic states~\cite{Wan2011,Burkov2011,Young2012,Wang2012,Weng2015,Zeng2015,TaS,MoC}. 
These topological semimetals can be classified according to the particular details of their electronic band structure, 
and the Weyl semimetal (WSM) is one of the most extensively studied cases of these classes. 
In Weyl semimetals, the conduction and valence bands cross linearly in momentum space via idoubly degenerated Weyl points, 
which behave as monopoles of Berry curvature with positive or negative chirality. 
In order to avoid the divergence of Berry curvature in momentum-space, 
the Weyl points must appear in pairs with opposite chiralities and the only way to annihilate this pair of Weyl points is to move them to the same $k$-point. 
Because of the existence of these Weyl points, WSMs can host several exotic transport properties in bulk, 
such as the chiral anomaly effect~\cite{Huang2015anomaly,Zhang2016ABJ,Wang2015NbP,Niemann2017}, 
gravitational anomaly effect~\cite{Gooth2017}, strong intrinsic anomalous Hall and spin Hall effects~\cite{Burkov:2011de,Xu2011,Sun2016}, 
and exhibiting a large magnetoresistance ~\cite{Shekhar2015,Ghimire2015NbAs,Huang2015anomaly,Zhang2016ABJ,Wang2015NbP,Luo2015,Moll2015}. 
Moreover, like TIs, WSMs also possess topologically protected surface states. 
Owing to the net Berry flux between Weyl points with opposite chiralities, 
the surface states in WSMs present as non-closed Fermi arcs terminating at two opposite Weyl points~\cite{Wan2011}.
Such Fermi arcs are different from the surface states in TIs and other topological materials, 
where the Fermi surfaces (FSs) present as closed curves at a fixed energy. 
As a result, the Fermi arc state plays the role of a fingerprint for WSMs, 
and provides the most direct way to verify the existence of Weyl points that can be measured by surface detection techniques, 
such as angle-resolved photoemission spectroscopy (ARPES) and scanning tunnelling microscopy (STM).
 
\begin{figure*}[htb]
\centering
\includegraphics[width=0.9\textwidth]{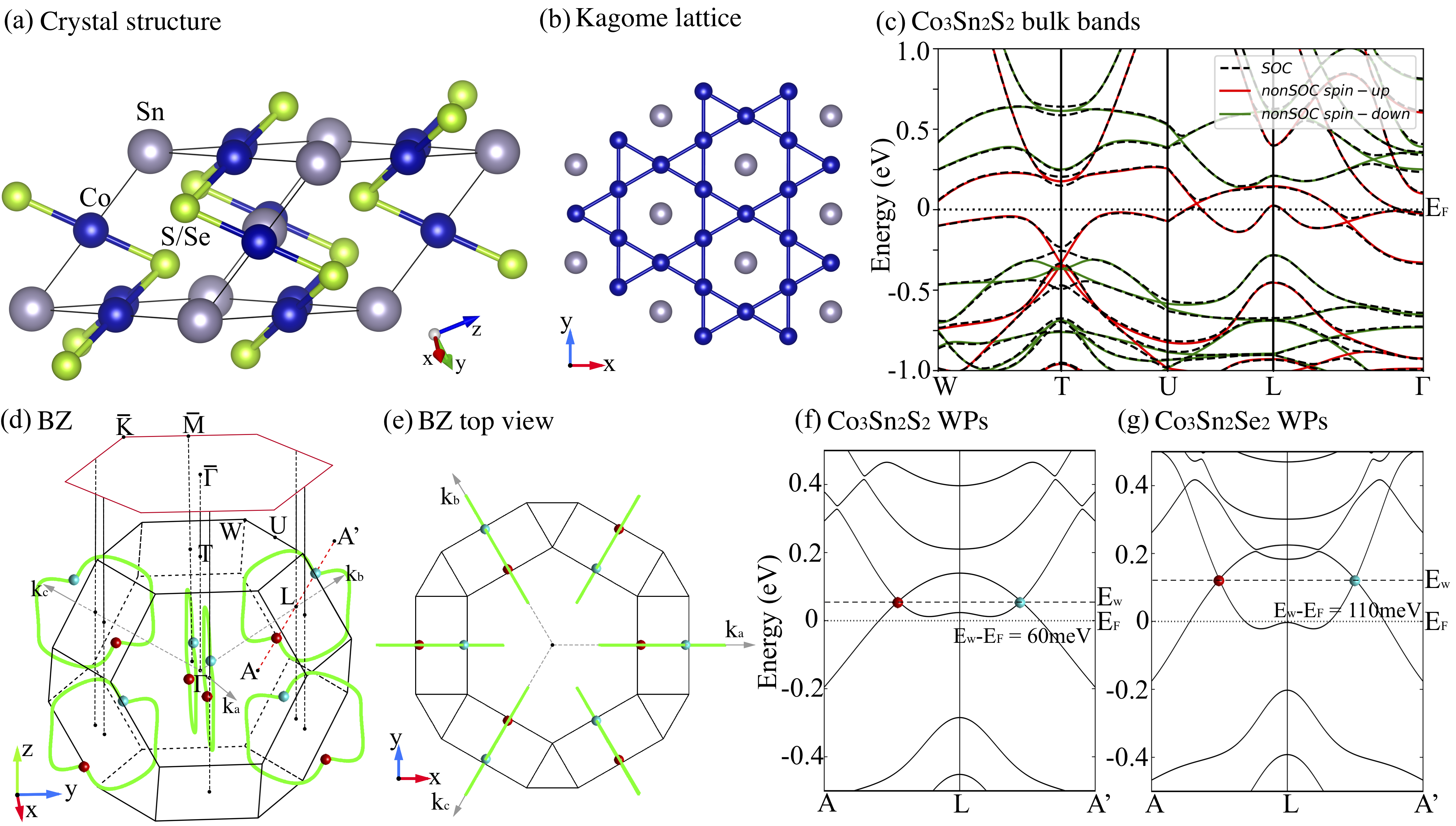}
   \caption{
(Color online) (a) Rhombohedral lattice structure of Co$_3$Sn$_2$S$_2$ and Co$_3$Sn$_2$Se$_2$, which exhibiting a layered lattice structure in the $xy$-plane. 
(b) The Kagome lattice structure of Co atoms in the $xy$-plane.
(c) Band structures of Co$_3$Sn$_2$S$_2$, with (black dashed lines) and without (red solid lines for spin up and green solid lines for spin down) SOC. 
(d) Location of the Weyl points (red and blue represent opposite chiralities) and nodal lines (green) in the 3D Brillouin zone (BZ) and the 2D BZ projected in the (001) direction. 
(e) Top view along z-direction of BZ with nodal lines and Weyl points.
(f) and (g) are energy dispersion along the $k$-path crossing a pair of Weyl points located on the same nodal line for Co$_3$Sn$_2$S$_2$ and Co$_3$Sn$_2$Se$_2$.
The Fermi level and energy of the Weyl points are labeled as $E_F$ and $E_W$, respectively.
}
\label{fig:lattice}
\end{figure*}

The existence of Weyl points need lifting the spin degeneracy of the system by breaking the inversion or time-reversal symmetry (or both). 
To date, many WSMs with the broken inversion symmetry have been theoretically predicted, 
and some have been experimentally verified, 
including the type-I WSM state in non-centrosymmetric transition-metal monophosphides (Ta/Nb)(As/P)~\cite{Xu2015TaAs,Lv2015TaAs,Yang2015TaAs,Liu2016NbPTaP,Xu2015NbAs,Belopolski2016NbP,Xu2016TaP,Souma2015NbP,Inoue2016, Batabyal2016, Zheng2016,Huang2015anomaly,Zhang2016ABJ,Wang2015NbP,Niemann2017},
and the type-II WSM state in WTe$_2$ and MoTe$_2$~\cite{Soluyanov2015WTe2,Sun2015MoTe2,Deng2016,Jiang2016,Liang2016,Huang2016}. 
In contrast, there are only a few candidates for magnetic WSMs which break time reversal symmetry, 
such as Y$_2$Ir$_2$O$_7$~\cite{Wan2011}, HgCr$_2$Se$_4$~\cite{Xu2011}, 
Co-based magnetic Heusler compounds~\cite{Wang_Heusler_2016,Chang2016,Kubler2016}, 
and heterostructures of TIs doped with magnetic impurities~\cite{Burkov:2011de}. 
However, so far, none of these proposed magnetic WSMs have been verified in experiments. 
Some potential reasons may be that the Weyl points in most of these candidate materials are situated far from the Fermi level ($E_F$) 
or the charge carrier density arising from trivial FSs is very large, 
which make physical phenomena arising from the Weyl points difficult to observe.

Very recently, a new magnetic WSM was proposed in the layered half-metal Co$_3$Sn$_2$S$_2$~\cite{Enk_2017,Wang2017}, 
whose Weyl points were predicted to be situated only 60 meV from the $E_F$, 
while simultaneously exhibiting a low charge carrier density. 
As a consequence, the properties of this material that are dominated by the Weyl points should be very easy to detect. 
On account of the large Berry curvature arising from the Weyl points and nodal lines opened by spin orbit coupling (SOC), 
the intrinsic anomalous Hall conductivity (AHC) was predicted to be as large as 1100 S/cm, in full agreement with transport measurements. 
Moreover, due to the combination of the large AHC and low charge carrier density, 
the anomalous Hall angle (AHA) was shown to be capable of reaching up to 20\%, which was never observed before. 
Therefore, the bulk transport properties provide strong evidence for the existence of Weyl points in Co$_3$Sn$_2$S$_2$. 
In this work, we have tried to further understand its topological features from the surface point view through
theoretical calculations, which should be also helpful for the future surface states measurements.

%%%%%%%%%%%%%%%%%%%%%%%%%%%%%%%%%%%%%%%%%%%%%%%%%%%%%%%%%%%%%%%%%%%%%%%%%%%%%%%%
\section{Methods}
To investigate the electronic and magnetic structures, 
we applied the Vienna $ab$--$initio$ simulation package (VASP)~\cite{kresse1996} 
for the first-principles calculations based on density functional theory (DFT) 
and chose the generalized gradient approximation (GGA) of the Perdew--Burke--Ernzerhof (PBE)~\cite{perdew1996} as
the exchange-correlation potential. 
The cut-off energy was 400 eV and the $k$-mesh for self-consistent calculation was 10$\times$10$\times$10.
In order to calculate the surface states, we projected the Bloch wave function into maximally localized Wannier functions (MLWFs)~\cite{Mostofi2008}
derived from the Co-3$d$, Sn-5$p$ and Se-4$p$ orbitals, 
and constructed a tight-binding Hamiltonian from the MLWF overlap matrix. 
Therefore the surface states were calculated in a half-infinite boundary condition using the Green's function method~\cite{Sancho1984,Sancho1985}. 
Moreover, we also calculated quasiparticle interference (QPI) patterns based on the Fourier-transformed surface local density of states from Green's function. 
The lattice constants are experimentally verified in Co$_3$Sn$_2$S$_2$~\cite{Enk_2017} 
and only predicted by DFT calculations in Co$_3$Sn$_2$Se$_2$ according to Ref. [51].

\begin{figure*}[htb]
\centering
\includegraphics[width=1.0\textwidth]{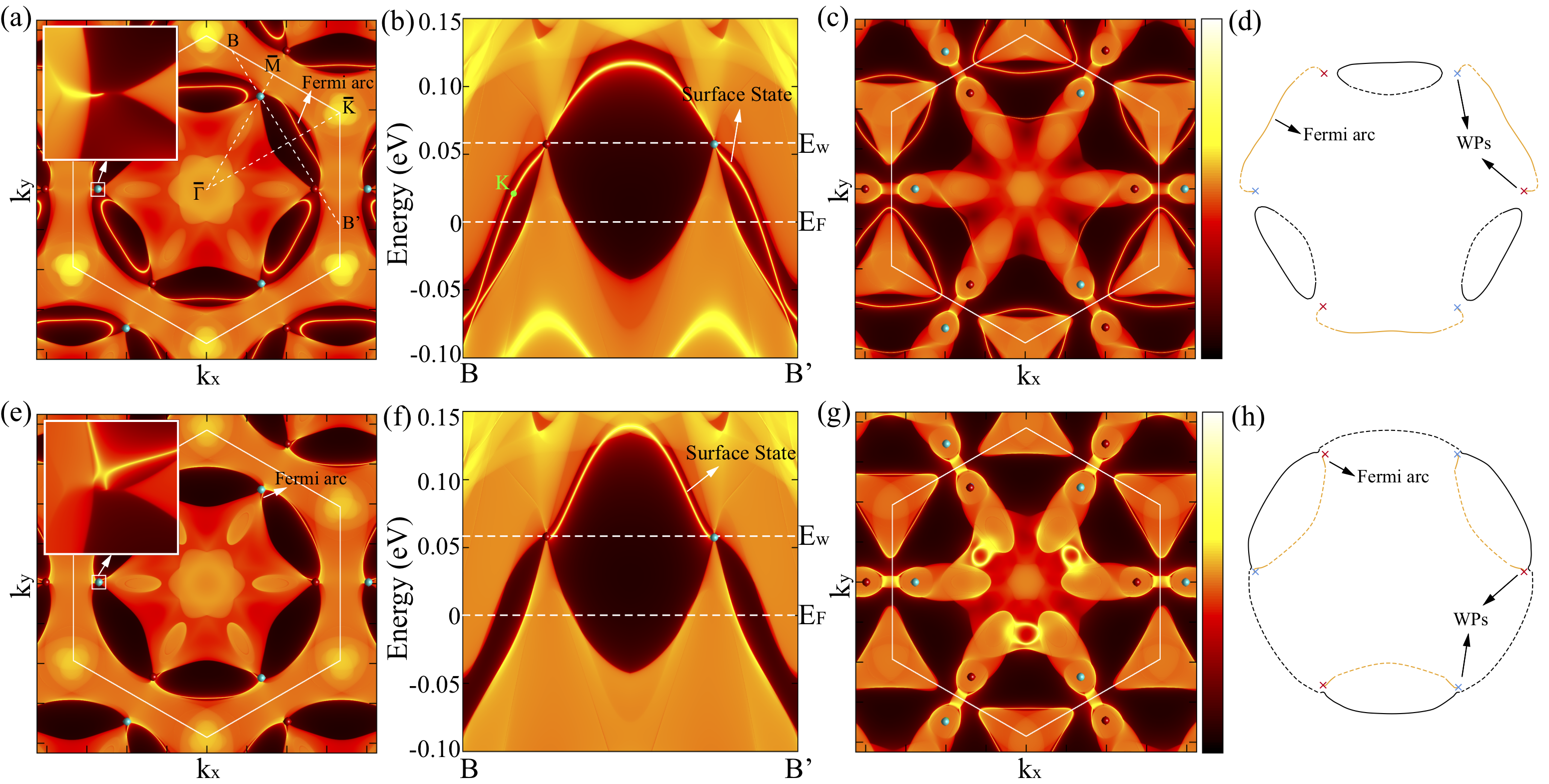}
   \caption{
(Color online) (001) surface states of Co$_3$Sn$_2$S$_2$ for S- and Sn-terminals. 
(a) Sn-terminated surface FS with energy fixed at the Weyl points.
(b) Energy dispersion for the Sn-terminated surface along a $k$-path crossing 
a pair of Weyl points connected by a Fermi arc. 
(c) Sn-terminated surface FS at the charge neutral point. 
(d) Schematic diagram of the FS distribution on the Sn-terminated surface. (e) S-terminated surface FS with energy fixed at the Weyl points. 
(f) Surface band structure for S-terminated states taken along the same direction as that in (b). 
(g) S-terminated surface FS at the Fermi energy, $E_F$. 
(h) Schematic diagram of the FS distribution on the S-terminated surface. 
The colour bar of (a)-(c) and (e)-(g) demonstrate the intensity of states.
The light (dark) colour demonstrates the occupied (unoccupied) states. 
The dashed curves in (d) and (h) represent surface states that merge into the bulk. 
The Fermi arcs and trivial FSs are shown in yellow and black, respectively.}
\label{fig:FS}
\end{figure*}

%%%%%%%%%%%%%%%%%%%%%%%%%%%%%%%%%%%%%%%%%%%%%%%%%%%%%%%%%%%%%%%%%%%%%%%%%%%%%%%%
\section{Results}
\begin{figure}[htb]
\centering
\includegraphics[width=0.48\textwidth]{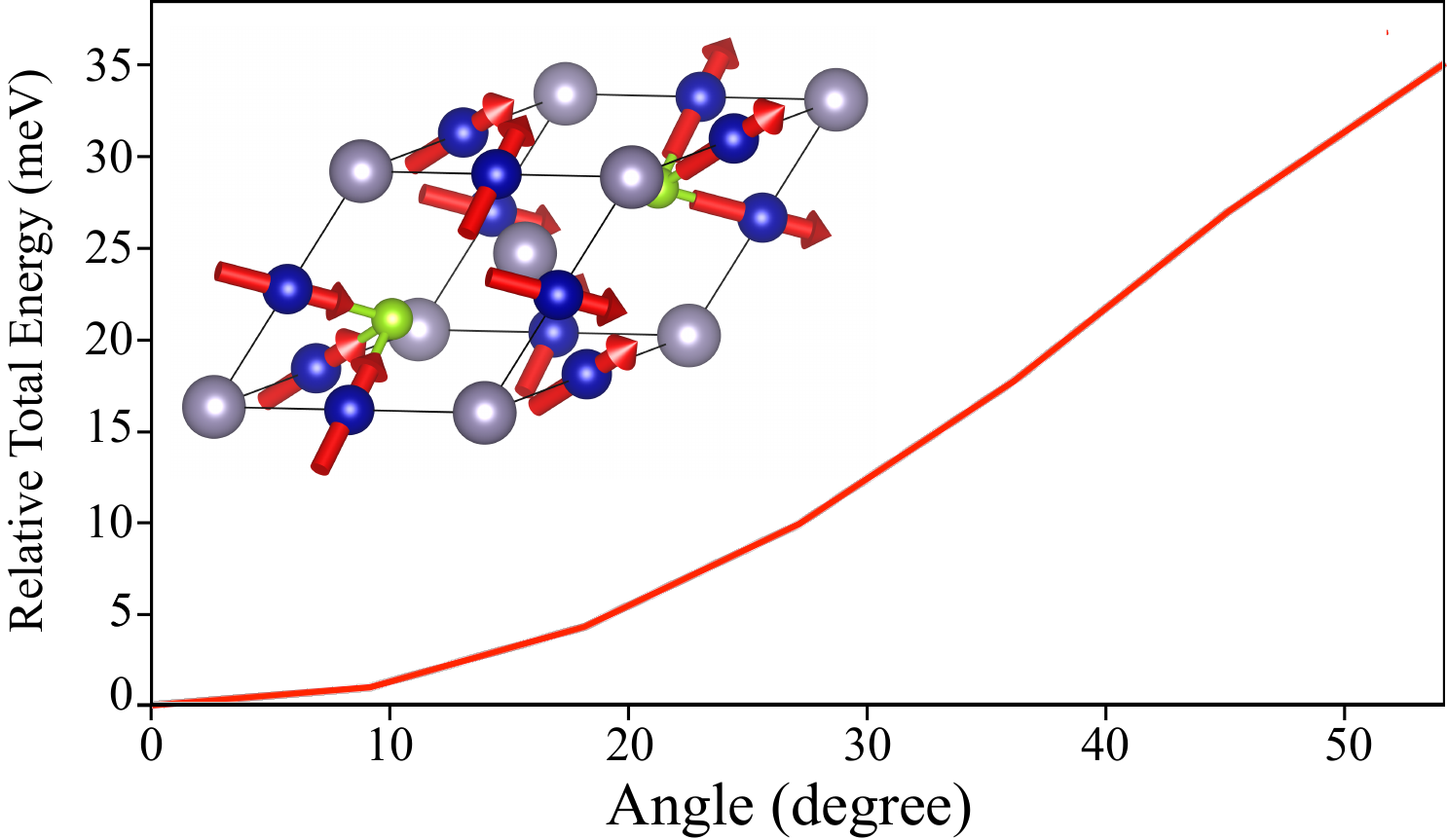}
   \caption{
(Color online) The change of the total energy with the increasing of the angle between z-axis and magnetic momenta of Co atoms.
When the magnetic momenta are parallel to the z-axis, it is most stable.}
\label{fig:mag}
\end{figure}

Co$_3$Sn$_2$S$_2$ and Co$_3$Sn$_2$Se$_2$ exhibit a rhombohedral lattice structure with 
the space group $R\bar{3}m$ (No. 166)~\cite{Sakai_2013,Weihrich_2015},
which has an inversion center, triple rotation axis and three mirror planes. 
The experimental lattice parameters of Co$_3$Sn$_2$S$_2$ (Co$_3$Sn$_2$Se$_2$) are 
a = 5.3757 $\AA$ (5.5635 $\AA$ ) and $\alpha$ = ${59.91647}^{\circ}$ (${58.33}^{\circ}$).
The Wyckoff site of atoms are Co [1/2, 0, 0], Sn1 [0, 0, 0], Sn2 [0.5, 0.5, 0.5] and S (Se) [n, n, n], where n = 0.2826 for S and n = 0.2864 for Se.
It can be viewed as a quasi-two-dimensional (2D) lattice stacked along the $z$-direction of the Cartesian coordinate system, 
as shown in Fig.~\ref{fig:lattice}(a). 
The magnetic Co atoms form a Kagome lattice in the $xy$-plane with their magnetic momentum aligned along the $z$-direction as shown in Fig.~\ref{fig:lattice}(b), 
while the Sn atoms assume the central positions of the hexagonal lattice structure. 
These Kagome lattice layers are sandwiched between neighbouring S or Se layers, 
and connected to each other through another Sn layer in the $z$-direction. 
So far, it is only known that Co$_3$Sn$_2$S$_2$ presents a net moment along $z$-direction.
Moreover, there are also some possibilities that each Co has a local in plane ($x-y$ plane) 
component. We have checked some possible magnetic structures with local momenta rotating
in the plane constructed by $z$-axis and Co-S bonding, and the $z$ component
of the local momenta along the positive direction. The comparison of them is
given in Fig.~\ref{fig:mag}, from which we can see that the total energy increases along
with the increasing of the angle between the local moments and $z$ axis.
Though there might be some other possibilities of the magnetic structure, we found that 
the existence of Weyl points and anomalous Hall effect are robust as long as the net      
magnetic moment is along $z$.

Since Co$_3$Sn$_2$S$_2$ and Co$_3$Sn$_2$Se$_2$ show very similar electronic band structure, and only 
Co$_3$Sn$_2$S$_2$ was successfully synthesised so far, we will focus on Co$_3$Sn$_2$S$_2$ as the example
for the detailed analysis. In the absence of SOC, Co$_3$Sn$_2$S$_2$ exhibits a half-metallic band structure, where the spin-up channel cuts $E_F$, 
while the spin-down channel is insulating with a band gap of 0.35 eV, as shown in Fig.~\ref{fig:lattice}(c). 
On account of this protection from reflection symmetry, 
the band inversion of the spin-up states results in linear band crossing in the form of nodal ring located at mirror plane.
Considering the inversion and $C_{3z}$ rotation symmetries, there are a total of six nodal rings in the whole Brillouin zone (BZ),
as indicated in Fig.~\ref{fig:lattice}(d, e).
The two linear band crossings indicated at $U$--$L$ and $L$--$\Gamma$ are just two single points of one nodal ring. 
Moreover, when SOC is taken into consideration, the nodal ring becomes gapped out, and one pair of linear crossings is preserved in the form 
of Weyl points with opposite topological charges of Chern numbers +1 and --1. 
At the same time, these Weyl points will not locate at the high symmetry k-path. 
The exact locations of one Weyl point for Co$_3$Sn$_2$S$_2$ and Co$_3$Sn$_2$Se$_2$ are about (0.313, --0.0866, --0.0866) and (--0.287, 0.106, 0.106) 
in fraction coordinate of primitive cell, and the other five Weyl points in the same BZ can also be obtained by inversion and $C_{3z}$ rotation symmetry.
In the end, the Weyl points of Co$_3$Sn$_2$S$_2$ and Co$_3$Sn$_2$Se$_2$ will lie 60 meV and 110 meV above $E_F$.
Hence, electron doping is preferred for detecting Weyl-point-dominated properties.

A typical feature for WSMs is the non-closed Fermi arc surface state.
For this quasi-2D lattice structure in the $xy$-plane, it is easily to obtain the (001) surface in experiments. 
So there are two possible cleaving planes, breaking the S--Co bonds or the Sn--S bonds. 
On account of the atomic separation in the $z$-direction, 
we found that the bonding between the Sn and S layers is much weaker, and this was confirmed by our total energy calculations. 
The formation energy required to break the Sn--S bonds is around 2.9 eV per cell.  
Therefore the Sn--S bonds are easier to cleave. 
Thus we get two different terminations for the cleaved (001) surface, i.e, S- and Sn-terminals, respectively.

\begin{figure}[htb]
\centering
\includegraphics[width=0.33\textwidth]{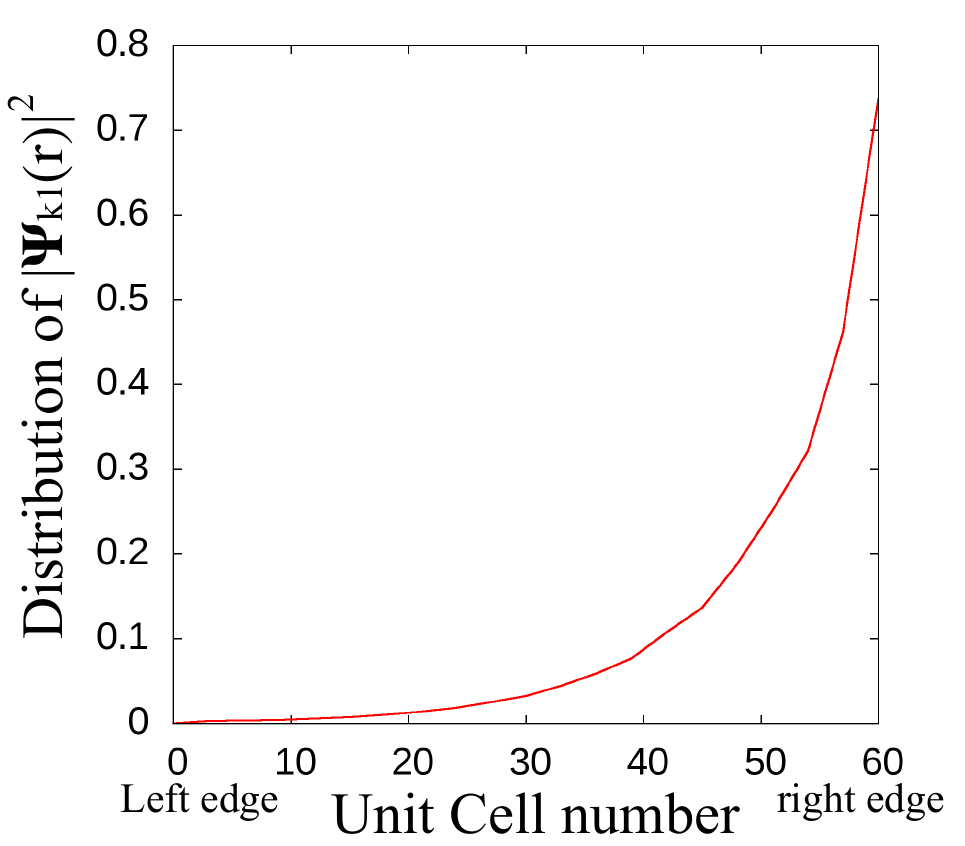}
   \caption{
(Color online) The decay of the surface state into the bulk. 
The wave function at the sample point K in Fig.~\ref{fig:FS}(b) on the surface state shows its decay into bulk as a function of the number of unit cell. The penetration depth is about 30-40 unit cells.
 }
\label{fig:decay}
\end{figure}

In order to check for possible Fermi arc states at $E_F$, we also analyzed the energy dispersion of the Fermi arc related states in energy space. 
As shown in Fig.~\ref{fig:FS}(b), the Fermi arc related bands can extend to $E_F$-0.1 eV from the energy of Weyl points ($E_W$), 
which offers good opportunity for detecting the Fermi arcs in the low energy region. 
Moreover, we also analyzed the decay of the surface state into the bulk.
Here, we pick the sample point $K$ on the surface state (Fig.~\ref{fig:FS}(b))              
to calculate the contribution of each unit cell in the slab model with thickness of 60 unit cells.
The result is shown in Fig.~\ref{fig:decay}, from which one can see that it needs at least 
30-40 unit cells to obtain the topological surface states.
At the same time, the corresponding FS is shown in Fig.~\ref{fig:FS}(c), 
where the Fermi arcs are located around the corners of the BZ, with the $\bar{K}$-points situated at the center of the three Fermi arcs. 
For the S-terminated state, most of the Fermi arcs merge into the bulk, making the visible Fermi arc states much shorter than those exhibited for the Sn-terminated states, 
see Fig.~\ref{fig:FS}(e, h). Moreover, the Fermi arc related bands extend to the higher energy region above the Weyl points, 
as indicated in Fig.~\ref{fig:FS}(f). Therefore, to observe the Fermi arcs from the S-terminal by ARPES, 
an electron-doped sample and surface electron-doping is necessary. 
However, we note that although potassium doping can be used to achieve small upward shifts in the surface chemical potential~\cite{Hossain2008}, 
this method does not work well in every situation.

\begin{figure*}[htb]
\centering
\includegraphics[width=0.8\textwidth]{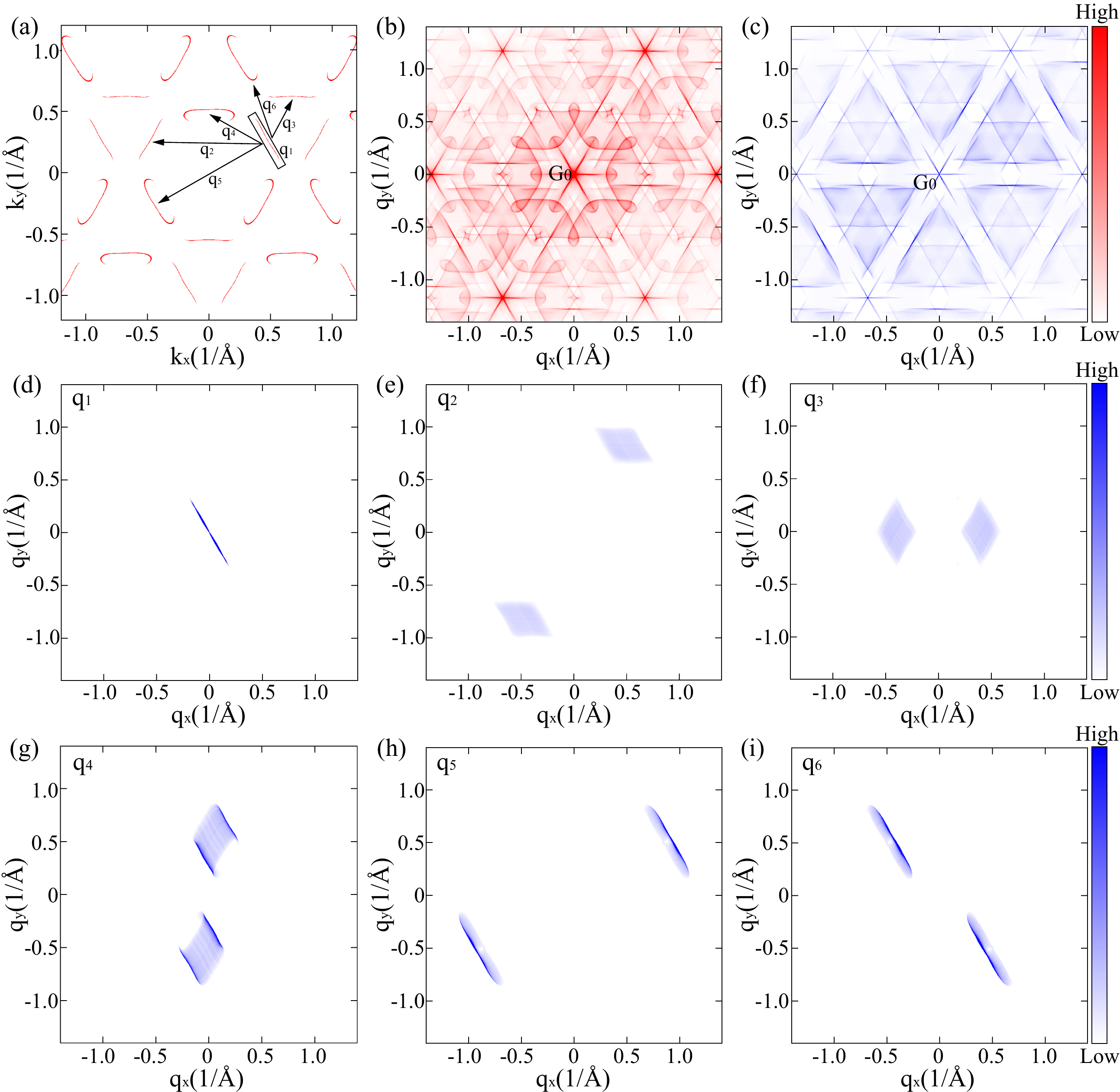}
   \caption{
(Color online) QPI patterns for an Sn-terminated surface.
(a) Surface FSs without those that merge into the bulk states. The independent scattering vectors $q_1$--$q_6$ are labeled with arrows.
(b) QPI patterns taking contributions from all possible scatterings into account. 
(c) QPI patterns only considering Fermi arc related scatterings.
(d) the intra--arc scattering of nontrivial Fermi arc.
(e)--(f) Specified inter--arc scattering patterns between two nontrivial Fermi arcs. 
(g)--(i) Details of the inter--arc scattering between nontrivial Fermi arc and trivial FS.
}
\label{fig:Sn}
\end{figure*}

Fig.~\ref{fig:FS} shows the surface FSs and energy dispersion corresponding to the two differently terminated surfaces. 
Fixing the energy at the Weyl points for the Sn-terminated surface, 
one can easily find the single surface FS starting from one Weyl point to another Weyl point with the opposite chirality, 
which is the typical expected behavior of Fermi arcs.
Taking the $C_{3z}$ rotation symmetry into account, there are three such Fermi arcs in the first BZ, as indicated in Fig.~\ref{fig:FS}(a, d). 
Although a little part of the Fermi arc merging into the bulk, the Fermi arc still extends to about 25\% of the reciprocal lattice vector, 
which is sufficiently long to be detected by an ARPES measurement.
  
\begin{figure*}[htb]
\centering
\includegraphics[width=0.8\textwidth]{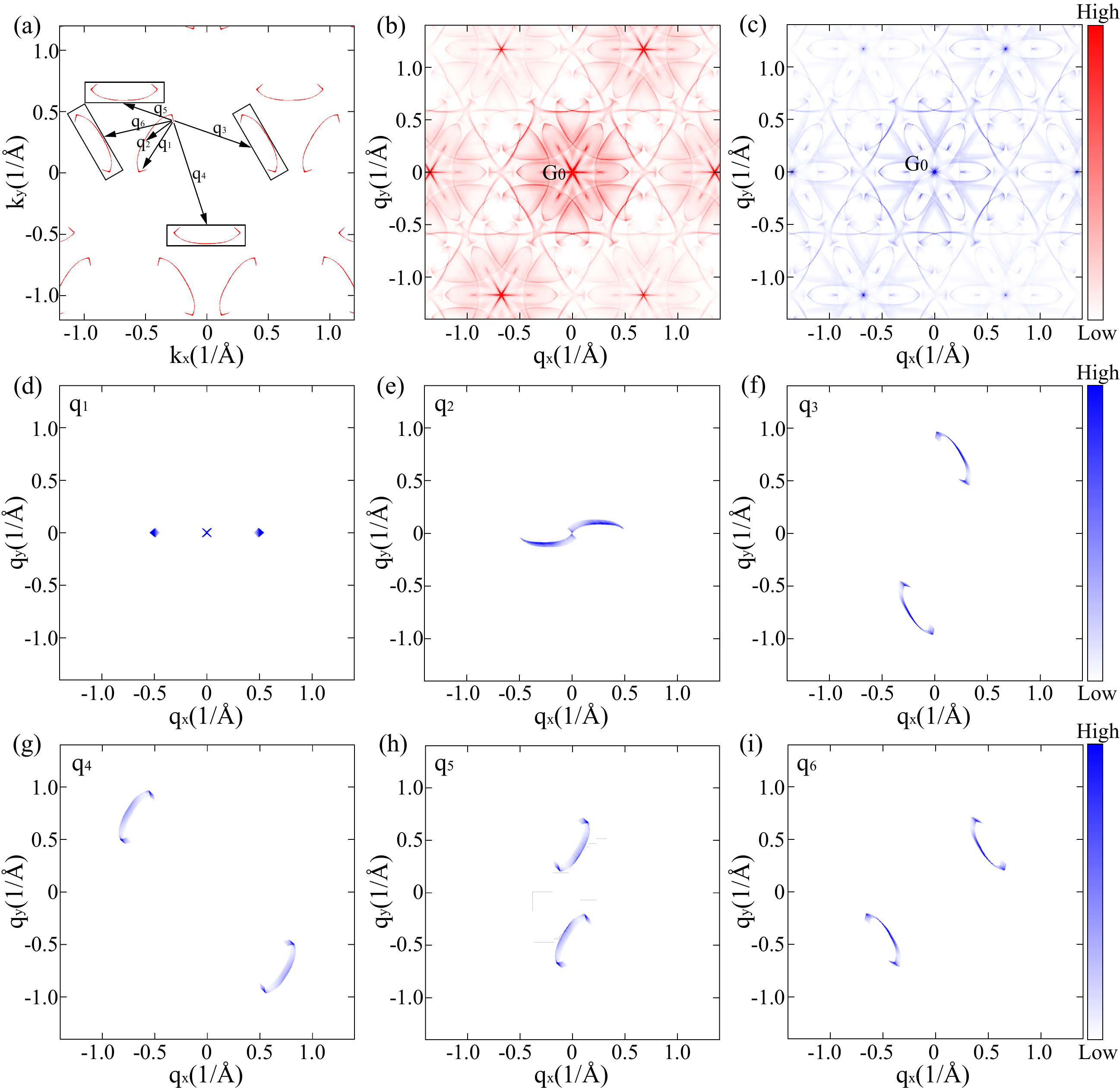}
   \caption{
(Color online) QPI patterns for an S-terminated surface.
(a) Surface FSs excluding parts that merge into the bulk states. The six Fermi arc related independent scattering vectors are labeled as $q_1$--$q_6$.
(b) Full QPI pattern taking all possible surface state scatterings channels into account.
(c) QPI pattern only considering the scattering from the nontrivial Fermi arcs. 
(d) Intensity of the $q_1$ arc-arc scattering.
(e)--(i) Intensity distributions of the five independent Fermi--arc--related inter--arc scattering $q_2$--$q_6$.
}
\label{fig:S}
\end{figure*}

Moreover, in a significant difference to WSMs that exhibit time-reversal symmetry, 
the surface state measurements of Co$_3$Sn$_2$S$_2$ need to consider the effects of magnetic domains. 
The size of these domains are typically around the order of micrometer, which is relatively small in terms of ARPES measurement. 
One solution to this problem is to align the magnetic moments along a particular direction by applying an external field. 
However, this is not generally feasible for ARPES measurements. Alternatively, the surface sensitive STM probe can gather enough information within the range of one domain,
provided that the surface is sufficiently smooth, since the sample surface can be scanned atom by atom. 
Furthermore, STM measurement yields another advantage by allowing us to measure the surface states at energies above $E_F$. 
Since the magnitude of the d$I$/d$V$ spectrum is proportional to the surface local density of states, 
the intensity of elastically scattering of electrons from surface defects can reveal the resulting QPI pattern.
The quantum interference between initial and final states ($k_i$ and $k_f$) at a constant energy results in a standard wave pattern indexed by a vector $q = k_f - k_i$.

In order to simulate such an STM measurement, we calculated the QPI from the
surface local density of states~\cite{Hosur2012, Hofmann2013,Kourtis2016}. 
In contrast to ARPES measurements, the QPI in STM is much more sensitive to the surface states, where the effects of $k_z$ are almost negligible. 
In this sense, we can justify the exclusion of the bulk projected states in our QPI calculations. 
Moreover, owing to the magnetic polarization in Co$_3$Sn$_2$S$_2$, 
the spins will be oriented along the $z$-direction with small in-plane components, 
and the QPI pattern is thus decided by the joint density of states (JDOS) without considering further spin selection rules. 

Fig.~\ref{fig:Sn} shows the resulting QPI pattern for the Sn-terminal with an energy
fixed at the Weyl points. In order to accurately simulate the scattering,
we only considered the states arising from the surface, as indicated in Fig.~\ref{fig:Sn}(a). 
Besides the Fermi arcs, another trivial FS states arised
from the dangling bond also exist, but they partly merge into the bulk, which appears
alternately with the Fermi arc around the $k_z$-axis. According to the
symmetry of the FSs, there are thus six independent scattering vectors
for each Fermi arc, three arising from arc--arc scattering and the
other three arising from scattering between the Fermi arcs and the trivial FS, which is shown in Fig.~\ref{fig:Sn}(a). 

Further details regarding the contributions of the different scattering intensities are presented in Fig.~\ref{fig:Sn}(d--i). 
As highlighted by the rectangle in Fig.~\ref{fig:Sn}(a), the Fermi arc appears as a nearly straight line at $E_F$. 
As a consequence, the intensity of the JDOS from intra--arc scattering forms the $G_0$ point centred line, 
which is indicated by $q_1$ in Fig.~\ref{fig:Sn}(d). 
In contrast, the JDOS patterns arising from scatterings between the Fermi arcs and trivial FSs are relatively wide due to the curvature of the trivial FS, 
as shown in Fig.~\ref{fig:Sn}(g--i), with the corresponding wave vectors labeled as $q_4$, $q_5$, and $q_6$.
To distinguish arc-related scatterings from the full QPI, we combined all the Fermi arc related JDOS in
Fig.~\ref{fig:Sn}(c), where includes the intra--arc scattering of Fermi arcs, 
the inter--arc scattering between two Fermi arcs, and the inter-arc scattering between Fermi arcs and trivial FSs.
Comparing with the full QPI, the arc-related QPI mainly dominate the six triangular regions near the central point $G_0$.

In contrast, the Fermi arc is very short for the S-terminated surface due to strong mixing between its central part with the bulk states. 
Fig.~\ref{fig:S}(a) displays the entire visible FSs, which include part of the Fermi arcs and part of the trivial FS that is not merged into the bulk. 
As a result of the short visible Fermi arc, arc--arc scattering only contributes a marginal amount to the overall QPI pattern. 
As indicated in Fig.~\ref{fig:S}(a, d), $q_1$, which represents arc--arc scattering from the same Fermi arc, 
yields only a tiny contribution to the whole $q$ space---one contribution located around the center of the BZ, 
and two others at ($q_x$, $q_y$)=($\pm0.5$, $q_y$).

Besides this arc--arc scattering, there are five other independent arc-related $q$ vectors, as indicated in Fig.~\ref{fig:S}(a). 
For convenience, we may view the two nearby broken Fermi arcs and trivial FSs as a single local FS due to the short distance between them, 
as marked by the black rectangles in Fig.~\ref{fig:S}(a). Then, using one visible apart of Fermi arc as the center, 
we are able to analyze other arc-related scattering channels, which can be classified into five groups. 
The indicated $q_2$ channel corresponds to the closest scattering between the Fermi arc and the trivial FS, 
which results in a $S$-shape pattern, as shown in Fig.~\ref{fig:S}(d). 
After taking the rotation and mirror symmetries into account,  
the resulting intensity pattern yields a $G_0$-centred pattern with three $\infty$-shape branches shown in Fig.~\ref{fig:S}(c). 
Besides the $\infty$-shape pattern, the intensity of other four scattering channels also form some non-closed 
patterns located relatively away from $G_0$, see Fig.~\ref{fig:S}(f--i). 
Considering all kinds of arc--related scattering, the QPI pattern shown in Fig.~\ref{fig:S}(c) can be obtained.

\section{Summary}
We theoretically studied the topological surface Fermi arc 
states in the magnetic Weyl semimetal Co$_3$Sn$_2$S$_2$ and 
Co$_3$Sn$_2$Se$_2$. The cleaved (001) surfaces have two possible 
terminations, Sn-terminal and S/Se-terminal, due to the weak 
chemical bonding between the Sn and S/Se layers. 
Fixing the energy at the Weyl points, the
 Fermi arcs are shown to be capable of extending to around 25\% 
of the reciprocal lattice vector in Sn-terminated states. 
Owing to the strong dispersion of the nontrivial surface state 
which extends from $E_F$--0.1eV to $E_F$+$E_W$, 
the large energy window provides good opportunity for the 
observation of the Fermi arc states in ARPES measurements. 
In contrast, the Fermi arcs for S/Se-terminated surfaces only exist above $E_F$ 
and are relatively short due to mixing of the surface and bulk states. 
Furthermore, we also simulated STM measurements of the differently terminated surfaces. 
These results will be helpful for a clear understanding of the 
surface Fermi arcs in these two magnetic WSMs.

This work was financially supported by the ERC Advanced Grant No. 291472 `Idea Heusler', 
ERC Advanced Grant No. 742068 `TOPMAT', and Deutsche Forschungsgemeinschaft DFG under 
SFB 1143. E.L. acknowledges support from the Alexander von Humboldt Foundation of Germany 
for his Fellowship and from the National Natural Science Foundation of China for his 
Excellent Young Scholarship (No. 51722106). L.M. would like to thank the MPI CPFS for 
its hospitality where part of the work was performed.

\bibliography{CoSnS_Arc.bib}

\end{document}